\documentclass{ws-ijbc}
\usepackage{ws-rotating}     
\usepackage {amsmath}
\usepackage {graphicx}
\usepackage {subfigure}
\begin{document}
\newcommand\ri{r_i}
\newcommand\rj{r_j}
\newcommand\vi{v_i}
\newcommand\gi{g_i}
\newcommand\mi{m_i}
\newcommand\mj{m_j}
\newcommand\Ji{J_i}
\newcommand\aji{g_{ji}}
\newcommand\aii{g_{ii}}
\newcommand\tdyn{t_{\rm dyn}}
\newcommand\rz{r_0}
\newcommand\Mt{M_{\rm tot}}
\newcommand\vtyp{v_{\rm typ}}
\newcommand\Ei{E_i}
\newcommand\Ki{K_i}
\newcommand\phii{\phi_{ii}}
\newcommand\phji{\phi_{ji}}
\newcommand\xv{{\bf x}}
\newcommand\yv{{\bf y}}

\catchline{}{}{}{}{} 

\markboth{Di Cintio \& Ciotti}{Relaxation with long-range forces}

\title{Relaxation of spherical systems with long-range interactions:\\
a numerical investigation}

\author{PIERFRANCESCO DI CINTIO}

\address{Department of Finite Systems, Max Planck Institute for the Physics 
of Complex Systems,\\ N\"othnitzer Stra\ss e 38, Dresden, D-01187 Germany\\
pdicint@pks.mpg.de}

\author{LUCA CIOTTI}
\address{Astronomy Department, Bologna University,\\
Via Ranzani 1, Bologna, 40127 Italy\\
luca.ciotti@unibo.it}

\maketitle

\begin{history}
\received{(to be inserted by publisher)}
\end{history}

\begin{abstract}
  The process of relaxation of a system of particles interacting with
  long-range forces is relevant to many areas of Physics. For obvious
  reasons, in Stellar Dynamics much attention has been paid to the
  case of $r^{-2}$ force law. However, recently the interest in
  alternative gravities emerged, and significant differences with
  respect to Newtonian gravity have been found in relaxation
  phenomena. Here we begin to explore this matter further, by using a
  numerical model of spherical shells interacting with an
  $r^{-\alpha}$ force law obeying the superposition principle.  We
  find that the virialization and phase-mixing times depend on the
  exponent $\alpha$, with small values of $\alpha$ corresponding to
  longer relaxation times, similarly to what happens when comparing for $N$-body
  simulations in classical gravity and in Modified Newtonian Dynamics.
\end{abstract}


\section{Introduction}

Phase Mixing and Violent Relaxation are important dynamical phenomena,
relevant in the astrophysical contest ([Lynden-Bell, 1967], [Binney \&
Tremaine, 2008]).  In fact $N$-body simulations revealed that the
final states of gravitating systems experiencing a collapse from {\it
  cold} and {\it clumpy} initial conditions are structurally very
similar to real elliptical galaxies ([van Albada, 1982], [Bertin {\it
  et al.}, 2005], [Nipoti {\it et al.}, 2006]). In addition, the
phase-space properties of the numerical end-products are remarkably
interesting, with a differential energy distribution described very
well by an exponential law, over a large value of energy values
([Binney, 1982], [Bertin \& Stiavelli, 1984], [Ciotti,
1991]). Finally, recent investigations of Modified Newtonian Dynamics
(MOND) showed that relaxation processes are much slower in MOND than
in Newtonian gravity ([Ciotti {\it et al.}, 2007], [Nipoti {\it et
  al.}, 2007]). This fact is relevant here, because in the weak limit
MOND forces behave qualitatively as $r^{-1}$.  Unfortunately, MOND is
a non-linear theory, so that numerical simulations are more difficult
than for systems with forces obeying the superposition principle.
With this paper we begin an exploration aimed at understanding how the
properties mentioned above depend on the specific nature of long-range
(power-law) interactions.  In particular, we study the relaxation of
systems governed by additive $r^{-\alpha}$ forces, focusing on the
simplified case of spherical systems.

\section{The model}

Following previous studies ([H\'enon, 1962], [Takizawa \& Inagaki,
1997], [Sanders, 2008], [Malekjiani {\it et al.}, 2009]), we
numerically integrate the motion of $N=10^4$ massive spherical shells.
We assume that a surface element of each shell, of mass $\delta m$ and
at vector position $\yv$, produces an acceleration at the point $\xv$
given by
\begin{equation}
{\bf g}=-G\delta m {\xv-\yv\over ||\xv-\yv||^{\alpha +1}},
\end{equation}
where $G$ is the ``gravitational'' constant, and the force index
$\alpha$, for reasons of convergence, is restricted to $\alpha<3$; the
case of Newtonian gravity is obtained for $\alpha=2$, while
$\alpha=-1$ corresponds to the case of harmonic force.  The equations
of motion for the shell $i$, of mass $\mi$ and radius $\ri$, are
\begin{equation}
{d\ri\over dt}=\vi,\quad
{d\vi\over dt}=a_i\equiv {\Ji^2\over\ri^3}+\aii+\sum_{j\neq i=1}^N\aji ,
\label{2}
\end{equation}
where $\vi$ and $\Ji$ are the radial velocity and the (constant)
angular momentum per unit mass of the shell, and $a_i$ is its effective radial
acceleration.  The field per unit mass
acting on each surface element of the shell $i$ due to the shell $j$
at $\rj\neq\ri$ is
\begin{equation}
\aji=-{G\mj\over 4\ri^2\rj}\times
\begin{cases}
\displaystyle{{(\ri+\rj)^{3-\alpha}-|\ri-\rj|^{3-\alpha}\over 3-\alpha}- 
(\ri^2-\rj^2){(\ri+\rj)^{1-\alpha}-|\ri-\rj|^{1-\alpha}\over\alpha-1},}
\quad (\alpha\neq 1)\\
\displaystyle{2\ri\rj-(\ri^2-\rj^2)\ln{|\ri-\rj|\over \ri+\rj},}
\quad\quad\quad\quad\quad 
\quad\quad\quad\quad\quad 
\quad\quad\quad\quad\quad 
\quad\quad
(\alpha =1)
\end{cases}
\end{equation}
while  
\begin{equation}
\aii=-{2^{1-\alpha}\over 3-\alpha}{G\mi\over\ri^{\alpha}}
\end{equation}
is the self-field per unit mass acting on each surface element of the
shell $i$; for $\alpha=2$ the celebrated Newton theorems are
recovered.  Note that $\aii$ can be obtained from $\aji$ by imposing
$\mj=\mi$ and taking the limit $\rj\to\ri$; the same limit is required
to obtain the force in case of shell superposition.

We performed experiments with different values of the exponent
$\alpha$, considering both radial collapses ($\Ji=0$) and collapses
with $\Ji\neq 0$. Numerically, when a shell collapses to
the origin, the sign of the velocity $\vi$ reverses to positive.  As
indicators of dynamical relaxation we consider the virial ratio
$2K/|W|$, the differential energy distribution $n(E)$, and the
phase-space evolution of the pairs $(\ri,\vi)$. We evolved 
the systems up to 50 dynamical times ($\tdyn$), where on dimensional
grounds 
\begin{equation}
\tdyn\equiv\sqrt{{2\rz^{\alpha+1}\over G\Mt}};
\end{equation}
$\Mt=\sum_i\mi$ is the total mass of the system, and $\rz$ is the
half-mass radius at $t=0$.  We explored different radial
distributions for the initial density profile, noticing that the final
products are not strongly dependent on it. The initial conditions are
characterized by a vanishing virial ratio, typical of cold collapses.
The integration scheme ([Di Cintio, 2009]) uses a standard 2$^{th}$
order leapfrog algorithm with constant timestep (e.g., see
[Grubm\"uller {\it et al.}, 1991]).

\section{The virial ratio}

We first consider the time evolution of the virial ratio $2K/|W|$.  As
the density distribution of the system can be written as
\begin{equation}
\rho(r)=\sum_{i=1}^N{\mi\over 4\pi\ri^2}\delta(r-\ri),
\end{equation}
the virial function $W$ \footnote{For potentials that are homogenous
  functions of $r$, as here for $\alpha\neq 1$, $W=(\alpha-1)U$, where
  $U$ is the potential energy. For $\alpha=1$ instead $W=-G\Mt^2/2$ is
  constant, as for deep-MOND forces [Nipoti {\it et al.}, 2007].}
([Binney \& Tremaine, 2008]) reduces to
\begin{equation}
W=4\pi\int_0^{\infty}\rho(r)\,g(r)\,r^3\,dr=\sum_{i=1}^N\mi\ri
\left(\aii+\sum_{j\neq i=1}^N\aji\right),
\end{equation}
while the total kinetic energy of the system is
\begin{equation}
K=\sum_{i=1}^N{\mi\over 2}\left(\vi^2+{\Ji^2\over\ri^2}\right).
\end{equation}
\begin{figure}[htb]
\begin{center}
\includegraphics[scale=0.45]{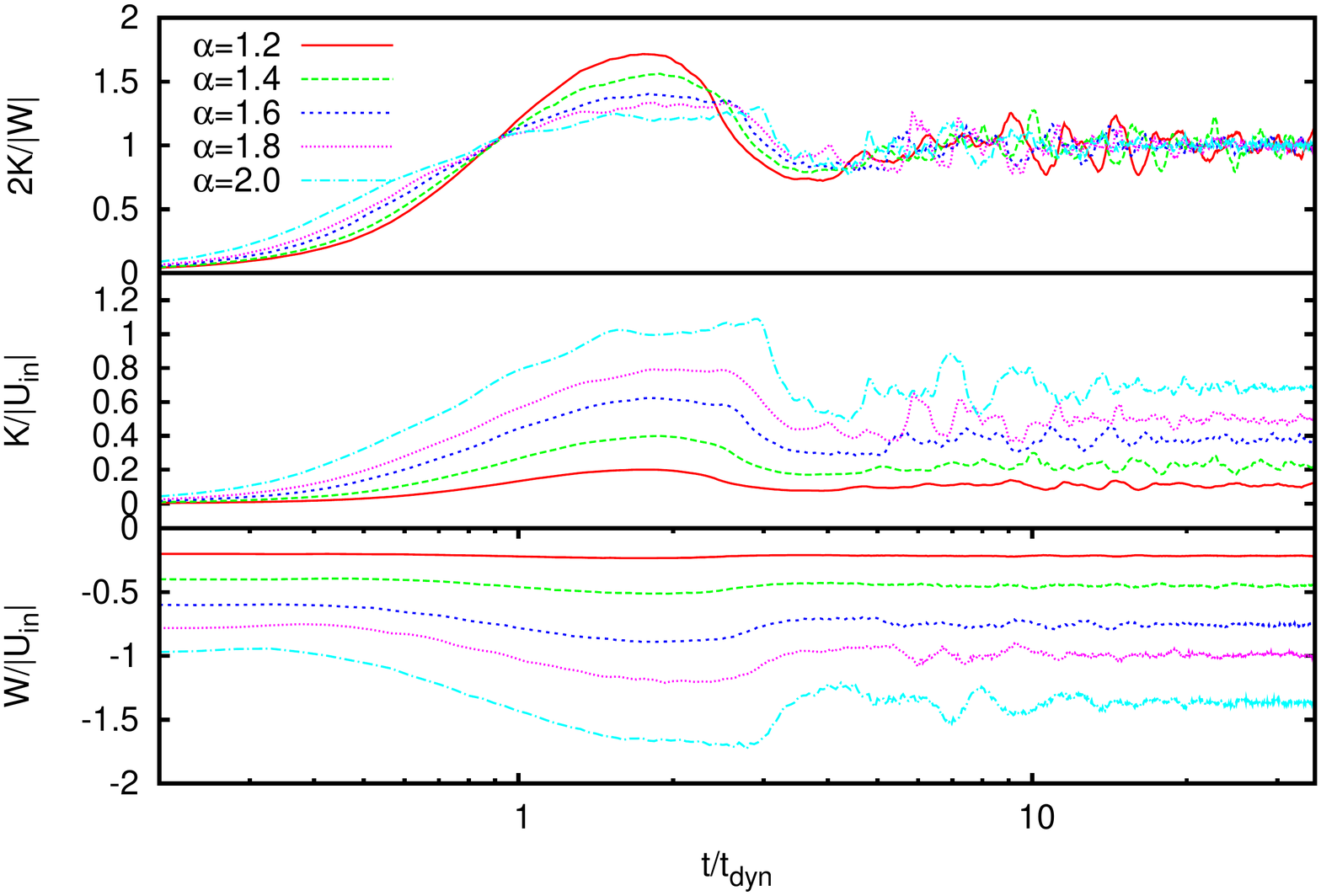} 
\end{center}
\caption{Top panel: evolution of the virial ratio over $\sim 36\tdyn$ for systems
  with $\Ji =0$, and different values of the force index $\alpha$
  (only bound shells have been considered in the plot). Low values of
  $\alpha$ correspond to larger and more long-lasting virial
  oscillations. Middle and bottom panel: kinetic energy and virial $W$
  in units of the initial binding energy $U_{\rm in}$. Virial theorem and energy
  conservation show that for cold collapses in absence of escapers, 
  $W=2(\alpha-1)U_{\rm in}/(3-\alpha)$ at virialization.}
\label{fig1}
\end{figure}
As apparent from Fig.~1, in purely radial collapses (and in collapses
with $\Ji\neq 0$, not shown), the evolution of the virial ratio shows
a violent jump followed by an oscillatory phase, around the
equilibrium value 1.  The violent phase lasts $\simeq 2\tdyn$, while
the subsequent virial oscillations last considerably longer. A certain
trend with $\alpha$ is apparent: as the exponent $\alpha$ increases
towards 2 (the force range shortens), the peak of the first and of the
successive jumps lowers. Remarkably, for the lowest value of $\alpha$
explored, (i.e., for the forces with the slowest decline with
distance), the virial oscillations continue for several tens of
dynamical times, similarly to what found in numerical simulations in
MOND collapses. We interpret this result as a clear trend of the
system behavior towards the harmonic oscillator case $\alpha=-1$, when
from eqs.~(3) and (4) it follows that $a_i=-G\Mt\ri+\Ji^2/\ri^3$, so
that each shell oscillates independently of the others and no energy
exchanges between shells can take place.  It is important to note that
the somewhat arbitrary definition of $\tdyn$ in eq.~(5) is found to be
a quite accurate measure of the true dynamical time, as the first peak
of the virial ratio occur at $t\simeq 2 \tdyn$ for all the explored
values of $\alpha$.

\section{Phase-space evolution}

The same behavior is also shown by phase-space evolution of the pairs
$(\ri,\vi)$, represented in Fig.~2 for systems with $\alpha=1.2$ and
$\alpha=2$. In the former case (top panels), the phase-space still
presents an ordered structure after $\simeq 10\tdyn$, symptom of a
less efficient mixing than in the Newtonian case (bottom panels).
This is reminiscent of the behavior of MOND systems, and the plots are
strikingly similar to those in Fig.~2 in [Ciotti {\it et al.}, 2007].
This finding is also confirmed by experiments with $\alpha \leq 1$ and
$\alpha>2$ (not shown). From Fig.~2 (bottom panels) is apparent how
shells are lost from the system with $\alpha=2$; note that ejection
is energetically impossible when $\alpha\leq 1$.

\begin{figure}[htb]
\begin{center}
\includegraphics[scale=0.47]{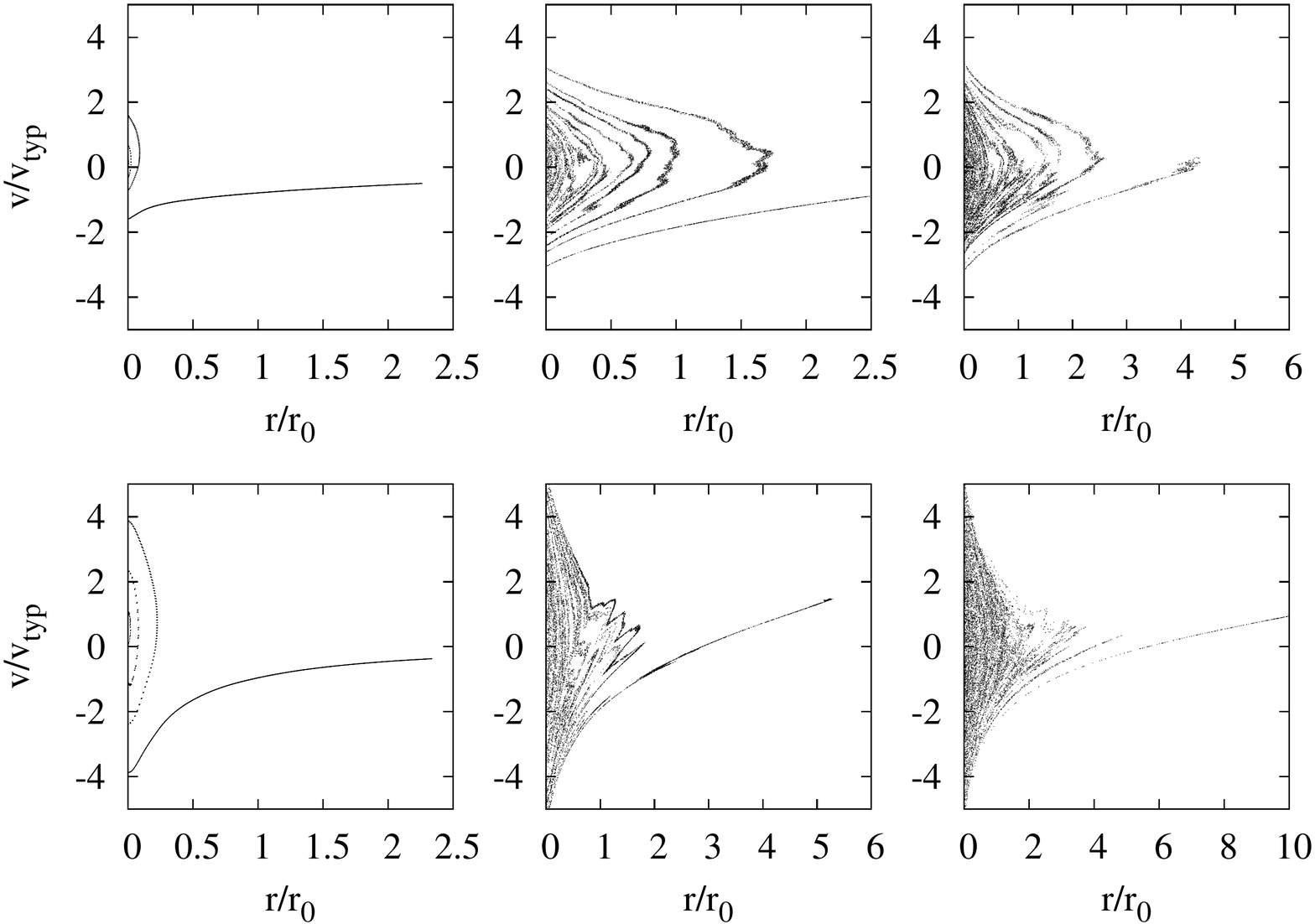} 
\end{center}
\caption{From left to the right: phase-space at $0.7\tdyn$, $7\tdyn$,
  and $14\tdyn$, for a system with $\alpha=1.2$ (top panels) and
  $\alpha=2$ (bottom panels). The scale velocity is
  $\vtyp\equiv\rz/\tdyn$, while $\rz$ and $\tdyn$ are defined
  in Sect.~2.  Note how phase mixing and shell ejection are more
  efficient in the case of Newtonian forces.}
\label{fig2}
\end{figure}
\begin{figure}[h!tb]
\begin{center}
\includegraphics[scale=0.45]{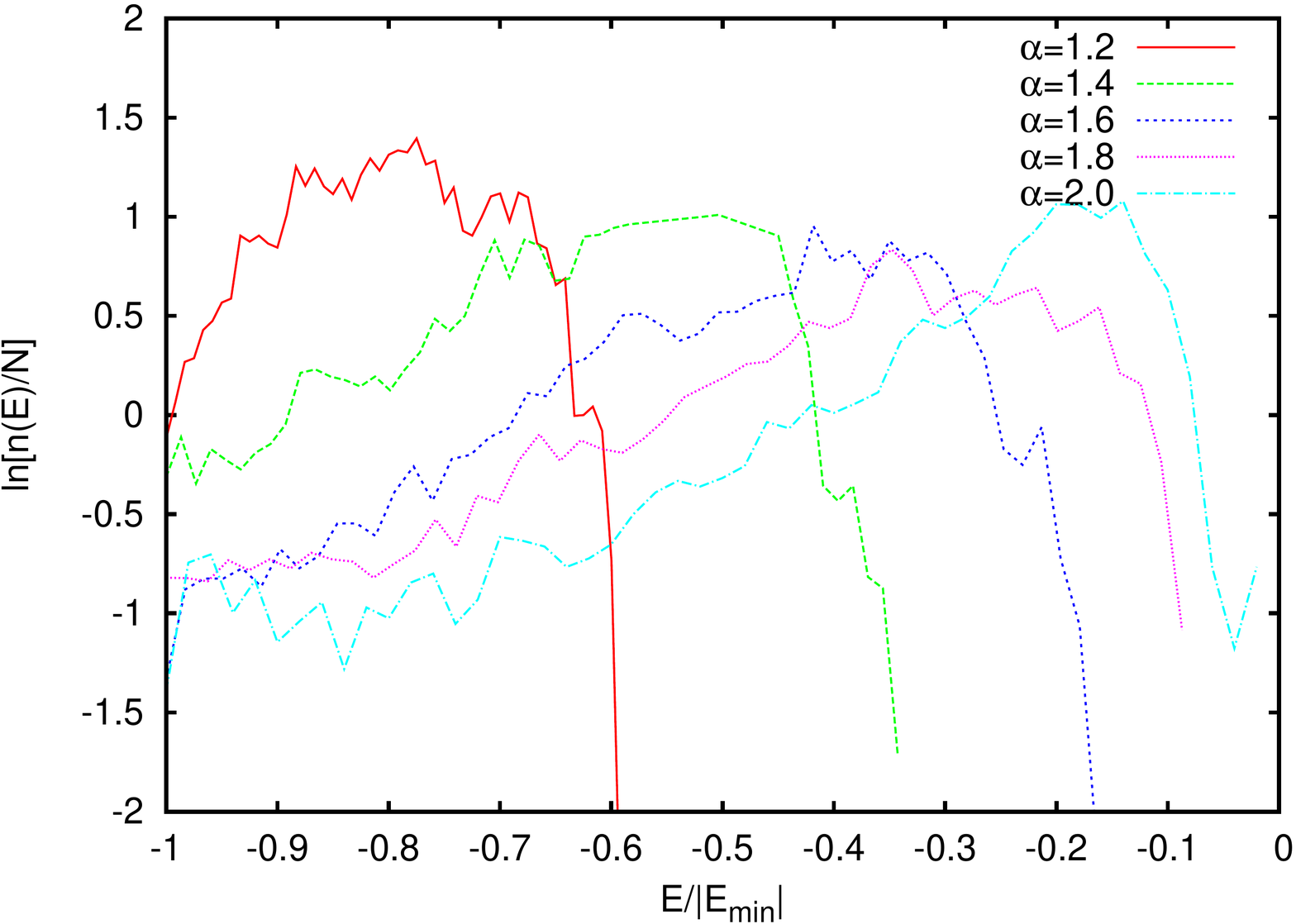} 
\end{center}
\caption{Final differential energy distribution, normalized to
  $E_{\rm min}/N$, as a function of $\alpha$. $E_{\rm min}$ is the
  minimum energy value attained by the shell distribution.}
\label{fig3}
\end{figure}

\section{Differential energy distribution}

We finally present the differential energy distribution at
virialization, $n(E)$. The quantity $n(E)dE$ is defined as the number
of shells with energy in the range $(E, E+dE)$. The total energy of
each shell is 
\begin{equation}
\Ei=\Ki + \mi\left({\phii\over 2} +\sum_{j \neq i=1}^N\phji\right),
\end{equation} 
where an integration gives the expressions for the
potential $\phji$ at $\ri$ due to the shell $j$, and the
self-potential\footnote{Note that $-d\phji/d\ri=\aji$, 
while  $-d\phii/d\ri=2\aii$. Note also that $\sum E_i$ is not the
total energy of the system.} $\phii$. For $\alpha\neq 1$
\begin{equation}
\label{due}
\phji=-{G\mj\over 2\ri\rj}
{(\ri+\rj)^{3-\alpha} -|\ri-\rj|^{3-\alpha}\over (\alpha-1)(3-\alpha)};\quad
\phii=-{G\mi 2^{2-\alpha}\over (\alpha-1)(3-\alpha)\ri^{\alpha-1}},
\end{equation}
while for $\alpha =1$
\begin{equation}
\label{due}
\phji={G\mj\over 4\ri\rj}
[(\ri+\rj)^2\ln (\ri+\rj)-(\ri-\rj)^2\ln |\ri-\rj| -2\ri\rj],\quad
\phii=G\mi\left (\ln\ri +\ln 2 -{1\over 2}\right).
\end{equation}
The formulae for the self-potential can be obtained by direct
integration or by taking $\mj=\mi$ and $\rj\to\ri$ in the expression
of $\phji$.  The final $n(E)$ for systems with $1.2<\alpha<2$ is shown
in Fig.~3. The distributions (albeit quite noisy, due to the limited
number of shells used in the simulations) roughly reproduce the
expected decline with increasing (absolute) energy predicted by an exponential
function
\begin{equation}
n(E)=n_0e^{-\beta |E|},\quad \beta>0,
\end{equation}
([Binney, 1982], [Ciotti, 1991]). However, the agreement of our results with eq.~(12), known to provide
an excellent fit over a large energy range of the virialized
end-products of $N$-body simulations of {\it cold} and {\it clumpy}
systems, is quite poor. This is not surprising, as the imposed
spherical symmetry of the shells does not allow the additional mixing
effects due to non-radial density inhomogeneities. We conclude by
noticing that the case $\alpha =1.2$ presents a curious,
non-monothonic trend with energy; we verified that this feature is not
a numerical artifact. We stress that also in MOND numerical
simulations (that should correspond, in a broad sense, to the
low-$\alpha$ cases) the final $n(E)$ is very different to that of
Newtonian simulations starting from the same initial conditions (see
Fig.~5 in Nipoti {\it et al.}, 2007).

\section{Discussion and conclusions}

With the aid of a simplified model of $N$ massive, spherically
symmetric and concentric shells, moving under the action of
long-range, power-law additive forces, we studied the relaxation
process leading to the final virialized states.  We find that for all
the explored values of $\alpha$ the relaxation process involves first
a rapid Violent Relaxation phase, followed by a longer and gentle
relaxation due to Phase Mixing.  For $\alpha$ increasing from values
$< 1$ to $>2$, the relaxation time (in units of dynamical time)
decreases steadily. In general, systems involving force laws with low
$\alpha$ have longer mixing times compared to those of systems with
high $\alpha$, that are by contrast characterized by faster
evaporation, resulting in a well marked core-halo segregation.  This
finding can be useful to improve our understanding of MOND, where the
relaxation mechanisms (due to the nonlinearity of the theory) are
still quite poorly understood; in fact, the cases of low $\alpha$
present an evolution similar to that of $N$-body MOND systems.
Finally, the $n(E)$ distribution of the final virialized states are
qualitatively similar (with one exception), increasing for increasing
energies, but are not described particulary well by an exponential law
(especially for low $\alpha$).  The next step in our study will be the
investigation of the relaxation of $N$-body systems with interparticle
additive forces as in eq.~(1). One of the main reasons for the
successive analysis is to abandon the assumption of spherical
symmetry, as it is expected that the increasing number of degrees of
freedom will produce a faster relaxation, and a better agreement with
an exponential $n(E)$.  Also, it will be interesting to investigate
how radial orbit instability, characteristic of self-gravitating
systems with most particles on radial orbits, depends on the specific
nature of the force law. Unfortunately, for generic values of
$\alpha$, a field equation for the potential (such as Poisson equation
for Newtonian gravity, or the $p$-Laplace equation for MOND systems)
does not exists, so that particle-mesh methods cannot be used, and
only the more time-expensive direct $N$-body numerical schemes are
available. An original code is currently under testing.

\nonumsection{Acknowledgements}\noindent This paper is dedicated to
the memory of dr. Giacomo Giampieri, with whom this project started.
We also thank the two Referees for useful comments and advice.
L.C. is supported by the MIUR grant PRIN2008.


\begin{thebibliography}{9}

\bibitem[Bertin \& Stiavelli (1984)]{bert84} Bertin, G. \& Stiavelli,
  M. [1984] ``Stellar dynamical models of elliptical systems'', {\it
    A\&A} {\bf 137}, 26-28.

\bibitem[Bertin {\it et al.} (2005)]{bert05} Bertin, G., Trenti, M. \&
  van Albada, T.S. [2005] ``A family of models of partially relaxed
  stellar systems. II. Comparison with the products of collisionless
  collapse'', {\it A\&A} {\bf 433}, 57-72.

\bibitem[Binney (1982)]{binney82} Binney, J. [1982] ``The phase space
  structure of $R^{1/4}$ galaxies - Are these galaxies 'isothermal'
  after all?'', {\it MNRAS} {\bf 200}, 951-964.

\bibitem[Binney \& Tremaine (2008)]{bt2008} Binney, J. \& Tremaine,
  S. [2008] ``Galactic Dynamics'', 2nd Ed. (Princeton University
  Press, USA).

\bibitem[Ciotti (1991)]{ciotti91} Ciotti, L. [1991] ``Stellar systems
  following the $R ^{1/m}$ luminosity law'', {\it A\&A} {\bf 249},
  99-106.

\bibitem[Ciotti {\it et al.} (2007)]{CNL07} Ciotti, L., Nipoti, C. \&
  Londrillo, P.  [2007] ``Phase-mixing in MOND'', {\it Proc.
    Int. Workshop on Collective phenomena in macroscopic systems}
  (Como, Italy) World Scientific pp.177-186.

\bibitem[Di Cintio (2009)]{dicintio2009} Di Cintio, P.F. [2009]
  ``Relaxation of dynamical systems with long-range interactions'',
  (Bologna University, Master Thesis), Unpublished.

\bibitem[Grubm\"uller {\it et al.} (1991)]{gm1991} Grubm\"uller, H.,
  Heller, H., Windemuth, A. \& Schulten, K. [1991] ``Generalized
  Verlet algorithm for efficient molecular dynamics simulations with
  long-range interaction'', {\it Mol. Sim.} {\bf 6}, 121-142.

\bibitem[Henon (1964)]{hen64} H\'enon, M. [1964] ``L'\'evolution
  initiale d'un amas sph\'erique'', {\it Annales d'Astrophysique} {\bf
    27}, 83-88.

\bibitem[Kandrup (1998)]{kan98} Kandrup, H.E. [1998] ``Violent
  Relaxation, Phase Mixing, and gravitational Landau damping'', {\it
    ApJ} {\bf 500}, 120-128.

\bibitem[Lynden-Bell (1967)]{LB67} Lynden-Bell, D. [1967]
  ``Statistical mechanics of violent relaxation in stellar systems'',
  {\it MNRAS} {\bf 136}, 101-121.

\bibitem[Malekjani {\it et al.} (2009)]{MRH09} Malekjani, M., Rahvar,
  S. \& Haghi, H. [2009] ``Spherical collapse in Modified Newtonian
  Dynamics'', {\it ApJ} {\bf 694}, 1220-1227.

\bibitem[Nipoti {\it et al.} (2006)]{NLC06} Nipoti, C., Londrillo,
  P. \& Ciotti, L. [2006] ``Dissipationless collapse, weak homology
  and central cores of elliptical galaxies'', {\it MNRAS} {\bf 370},
  681-960.

\bibitem[Nipoti {\it et al.} (2007)]{NLC07} Nipoti, C., Londrillo,
  P. \& Ciotti, L. [2007] ``Dissipationless collapses in Modified
  Newtonian Dynamics'', {\it ApJ} {\bf 660}, 256-266.

\bibitem[Sanders (2008)]{Sanders08} Sanders, R.H. [2008] ``Forming
 galaxies with MOND'', {\it MNRAS} {\bf 386}, 1588-1596.

\bibitem[Takizawa \& Inagaki (1997)]{Takizawa97} Takizawa, M. \&
  Inagaki, S. [1997] ``Violent relaxation of spherical stellar
  systems'', {\it astro-ph9702002}. 

\bibitem[van Albada (1982)]{vanAlbada82} van Albada, T.S. [1982]
  ``Dissipationless galaxy formation and the $R^{1/4}$ law'', {\it
    MNRAS} {\bf 201}, 939-955.

\end{thebibliography}
\end{document}